\documentclass[onecollarge,natbib]{svjour2}
\bibpunct{[}{]}{;}{n}{}{,} 
\smartqed  
\usepackage{graphicx}
\usepackage{amssymb,amsmath}
\usepackage{mathptmx}      
\def\bea{\begin{eqnarray}}
\def\eea{\end{eqnarray}}

\journalname{Few-Body Systems (EFB22)}
\begin{document}

\title{Partial Wave Analysis of Chiral NN Interactions  
~\thanks{
Supported by Spanish DGI (grant FIS2011-24149), Junta de
Andaluc\'{\i}a (grant FQM225) and the Mexican CONACYT.  
}
}


\author{R. Navarro P\'erez \and J. E. Amaro \and 
E. Ruiz Arriola    
}


\institute{
R. Navarro P\'erez \at
              Departamento
  de Fisica At\'omica, Molecular y Nuclear and Instituto Carlos I de
  Fisica Te\'orica y Computacional, Universidad de Granada, E-18071
  Granada, Spain              \email{rnavarrop@ugr.es}           
\and J. E. Amaro \at
              Departamento
  de Fisica At\'omica, Molecular y Nuclear and Instituto Carlos I de
  Fisica Te\'orica y Computacional, Universidad de Granada, E-18071
  Granada, Spain              \email{amaro@ugr.es}           
\and E. Ruiz Arriola \at
              Departamento
  de Fisica At\'omica, Molecular y Nuclear and Instituto Carlos I de
  Fisica Te\'orica y Computacional, Universidad de Granada, E-18071
  Granada, Spain              \email{earriola@ugr.es}           
} \date{Presented by R. N. P. at 22th European Conference On Few-Body
  Problems In Physics: EFB22 \\ 9 - 13 Sep 2013, Krakow (Poland)}

\maketitle

\begin{abstract}
We analyze chiral interactions to N2LO on the light of proton-proton
and neutron-proton scattering data published from 1950 till 2013 and
discuss conditions under which the chiral coefficients can be
extracted.  
\end{abstract}

\section{Introduction}
\label{intro}

While the NN interaction is traditionally acknowledged as a key
building block in Nuclear Physics, the possibility of describing it
using chiral symmetry and effective field theory methods has been a
fascinating pasttime for Nuclear theoreticians for more than 20 years
as it offers a link to the underlying quark and gluon dynamics of
QCD~(see e.g. \cite{Epelbaum:2008ga,Machleidt:2011zz} for reviews). A
crucial feature is the correct determination of the chiral constants
$c_1$, $c_3$ and $c_4$ which appear both in $\pi N$ as well as in $NN$
scattering as a TPE
contributions~\cite{Kaiser:1997mw,Rentmeester:1999vw}. Our purpuse is
to extract them with errors from the analysis of the about 8000
scattering data collected from 1950 till 2013. This is based in our
previous works~\cite{
  NavarroPerez:2011fm,NavarroPerez:2012qf,NavarroPerez:2012vr,
  Perez:2012kt, NavarroPerez:2012qr,Perez:2013za,Perez:2013mwa,Perez:2013jpa,Amaro:2013zka}.

\begin{figure*}
\centering
\includegraphics[width=0.3\linewidth,height=0.3\linewidth]{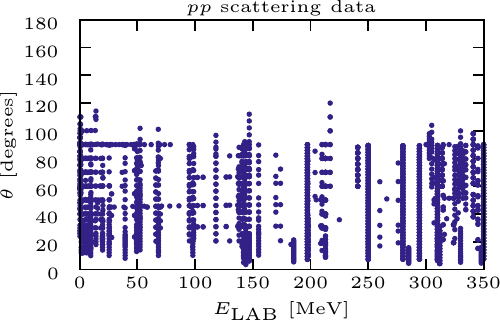}
\includegraphics[width=0.3\linewidth,height=0.3\linewidth]{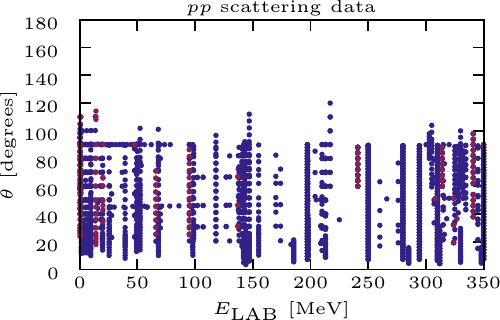}
\includegraphics[width=0.3\linewidth,height=0.3\linewidth]{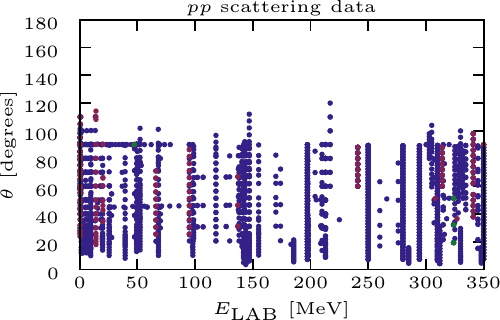}
\vskip.5cm 
 \includegraphics[width=0.3\linewidth,height=0.3\linewidth]{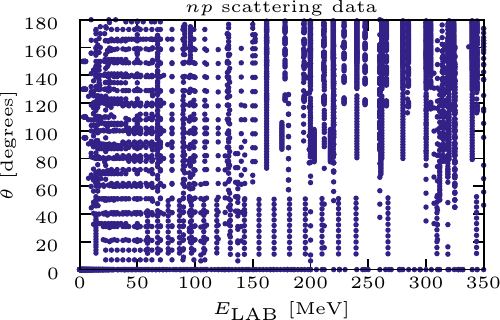}
\includegraphics[width=0.3\linewidth,height=0.3\linewidth]{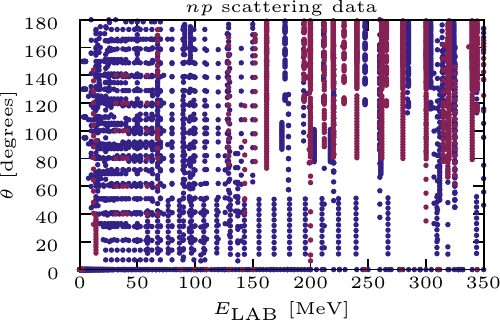}
\includegraphics[width=0.3\linewidth,height=0.3\linewidth]{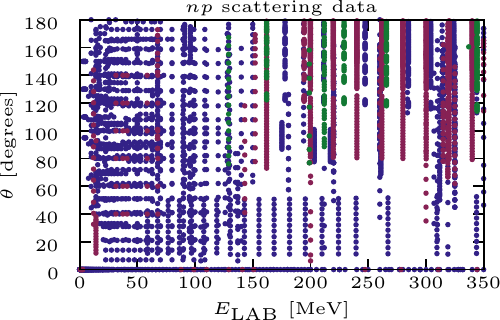}
\caption{Abundance plots for pp (top panel) and np (bottom panel)
  scattering data. Full data base (left panel). Standard $3\sigma$
  criterion (middle panel). Self-consistent $3\sigma$ criterion (right
  panel). We show accepted data (blue), rejected data (red) and
  recovered data (green).  }
\label{fig:NN-abundance}       
\end{figure*}

\begin{figure*}
\centering
 \includegraphics[width=8.5cm, height=6cm]{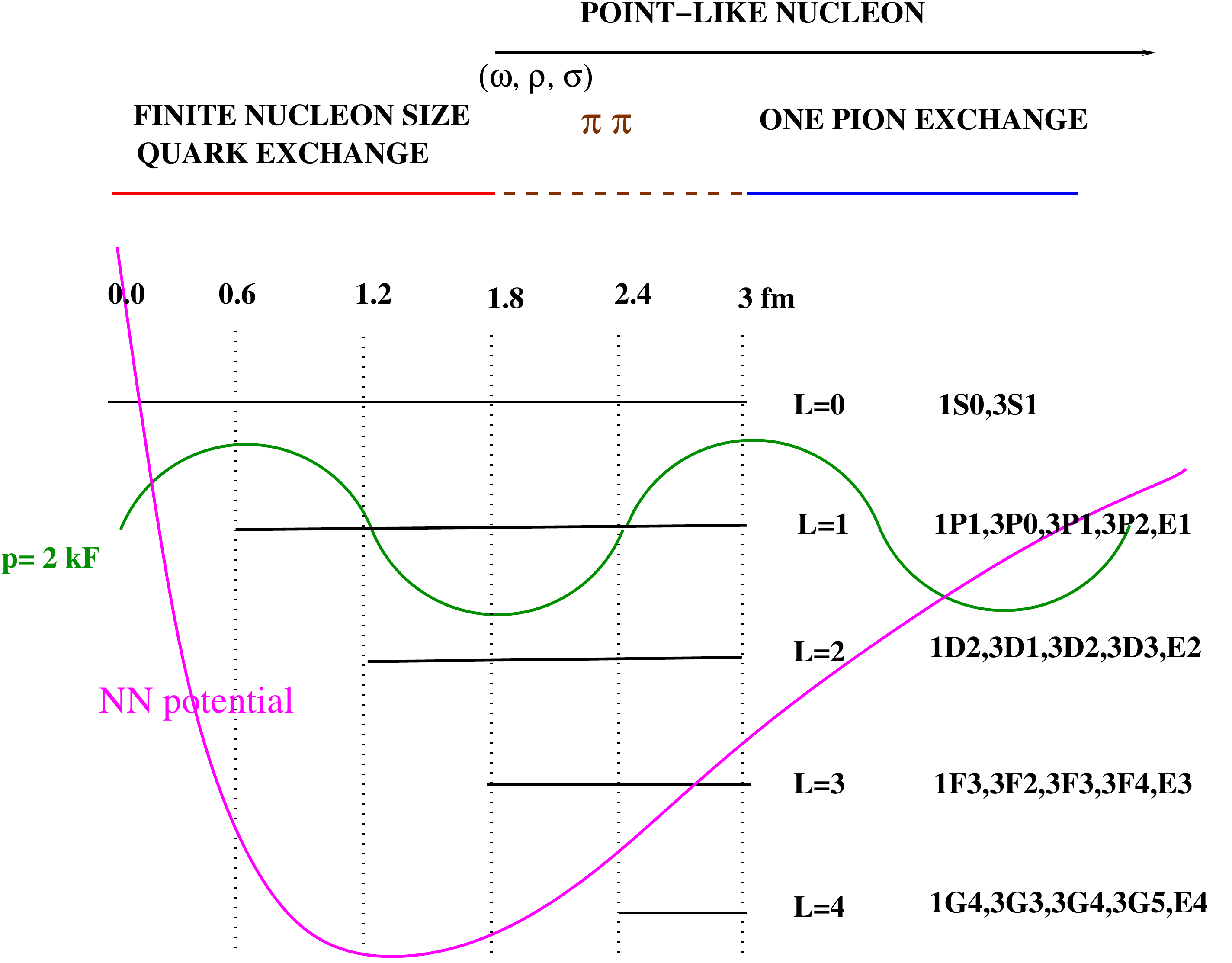}
\hskip2cm
  \includegraphics[width=0.25\textwidth]{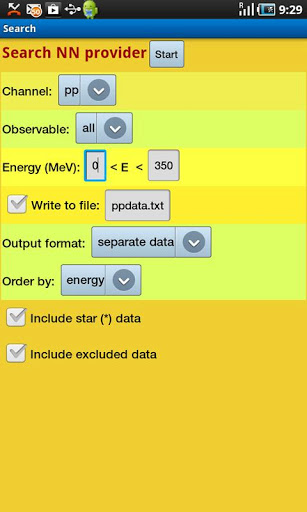}
\caption{{\bf Left panel:} Anatomy of the NN interaction showing the
  different regions as a function of the distance (in fm) for a
  resolution $\Delta r=0.6 {\rm fm}$ (see main text). {\bf Right panel:} The
  NN provider Android app, available at  Google Play Store.}
\label{fig:NN-anatomy}       
\end{figure*}

\section{Anatomy of the NN interaction and the number of fitting parameters}

In Fig.~\ref{fig:NN-abundance} we show the abundance plots for a total
number of 7709 pp and np data (the total number of 8124 fitting data
includes 415 normalization data provided by experimentalists), in the
LAB energy-angle plane.  Most high quality
fits~\cite{Stoks:1993tb,Stoks:1994wp,Wiringa:1994wb,Machleidt:2000ge,Gross:2008ps}
which have historically been capable of fitting their contemporary NN
scattering data with $\chi^2/{\rm d.o.f} \lesssim 1$ require about 40
parameters for the unknown part of the interaction. To understand the
rationale of this, the anatomy of the NN interaction below pion
production threshold is sketched in Fig.~\ref{fig:NN-anatomy}.  The
maximal CM momentum corresponding to the inelastic process $NN \to
NN\pi$, which is roughly $p_{\rm CM}= \sqrt{m_\pi M_N}$.  This
corresponds to a de Broglie wavelength, which we identify with the
shortest resolution scale $\Delta r\sim \hbar /p_{\rm max} \sim 0.6
{\rm fm}$.  For comparison we also depict a free spherical wave, $\sin
(p r)$ with $p=2 k_F$ relevant for nuclear matter.  The idea is to
coarse grain the interaction down to that scale.  On the other hand,
nucleons are composite and extended particles made of three quarks,
$p=uud$ and $n=udd$, thus we must distinguish between the overlapping
and non-overlapping regions as measured by the interaction. For
instance, the classical electrostatic interaction the pp potential at
a distance $\vec r$ would be
\begin{eqnarray}
V_{pp,{\rm EM}}(r) = \int d^3 \vec r_1 d^3 \vec r_2 \frac{\rho_p(\vec r_1)\rho_p(\vec r_2)}{|\vec r_1-\vec r_2 - \vec r|} 
= \int \frac{d^3  \vec q}{(2\pi)^3} \frac{4 \pi e^ 2}{\vec q^2} |G_{E,p}(\vec q)|^2 e^{i \vec q \cdot \vec r} 
\sim \frac{e^ 2}{r} \qquad r \ge 2 {\rm fm}
\end{eqnarray}
where $G_{E,p}(\vec q)$ is the proton electric form factor (we take a
dipole). Thus, regarding EM interaction the proton behaves as a
point-like particle for $r \ge 2 {\rm fm}$ since $V_{pp,{\rm E}} ( 2
{\rm fm})= 0.714 {\rm MeV}$ to be compared to the point-like value
$0.719{\rm MeV}$.  For np one has $V_{np,E} ( 2 {\rm fm})=0.001 {\rm
  MeV}$ compared to a vanishing point-like electric interaction. A
similar situation happens for the strong part of the
interaction. Using cluster chiral quark model
calculation~\cite{RuizArriola:2009vp,Cordon:2011yd} in the
Born-Oppenheimer approximation with finite nucleon and $N\Delta$
transition form factors, one sees point-like spin-flavour van der
Waals interactions with OPE and TPE {\it above} $2 {\rm
  fm}$. Likewise, one can also check that {\it
  above} $3 {\rm fm}$ the main contribution is just OPE. This is consistent
with the regularization used for OPE in high quality
potentials~\cite{Stoks:1993tb,Stoks:1994wp,Wiringa:1994wb}. If we
switch off this {\it known} piece, we are left with a {\it unknown}
potential with a finite range $r_c=3{\rm fm}$.  For such a truncated
potential, the maximal angular momentum needed for convergence of the
partial wave expansion is $ l_{\rm max} = p_{\rm max} r_c = r_c /
\Delta r = N $. On the other hand, the minimal distance where the
centrifugal barrier dominates corresponds to $l(l+1)/r_{\rm min}^2 \le
p^2 $, which is $r_{\rm rmin}=0.7,1.2,1.7,2.2,2.7 {\rm fm}$ for
$l=1,2,3,4,5$ respectively. Thus, for a given $l \le l_{\rm max}$ we
can count the number of points between $r_{\rm min}$ and $r_c$ sampled
at a resolution $\Delta r = 1/\sqrt{M_N m_\pi}$, see
Fig.~\ref{fig:NN-anatomy}, which means $l_{\rm max}= N= 3,4,5$ for
$r_c=1.8,2.4,3$. We count partial waves according to their threshold
behaviour in coupled channels~\cite{PavonValderrama:2005ku}, namely $
^{2S+1}L_J={\cal O}(p^{2L})$, $E_J= {\cal O}(p^{2J})$ with $E_J$ being
the tensor mixing waves.  The number of parameters for an unknown
interaction below $r_c> 2 {\rm fm}$ and momentum $p \le p_{\rm max}=2
{\rm fm}^{-1}$ becomes
\begin{eqnarray}
N_{\rm par} (r_c) \sim 60,38,21 \qquad r_c = 3,2.4,1.8 {\rm fm} 
\label{eq:Npar}
\end{eqnarray}
This counting argument does not consider that some parameters may be
either accidentally small or turn out to be compatible with zero. A
polynomial counting to order $\nu$ in momentum, gives a hermitian real
potential $V_{l',l}^{S,J}(p',p)= p'^{l´} p^l \sum_\nu \sum_k
c_{l',l}^{S,J,(N,k)}(p')^{\nu}p^{\nu-2k} $ with $N( {\cal
  O}(p^\nu))=2,7,19,41$ total number of $c_{l',l}^{S,J,(N,k)}$
parameters for $\nu=0,2,4,6$ respectively. The expansion has a
convergence radius of $|p'|,|p| < m_\pi/2$, which is extended to $n
m_\pi /2$ after additive inclusion of $n \pi$ exchange. Thus for $p
\lesssim 3m_\pi/2$ one needs $2\pi$ exchange and just 9
coefficients~\cite{Ekstrom:2013kea}. This corresponds to take $\Delta
r \sim 1 {\rm fm}$ and or $E_{\rm LAB} \sim 90 {\rm MeV}$.

\section{Delta-shell potential fits}

There remains the question on {\it how} to encode the {\it unknown}
part of the interaction which should be sampled, or coarse grained, at
least with $\Delta r$ resolution~\cite{NavarroPerez:2011fm}. Following
a remarkable and forgotten paper by Aviles~\cite{Aviles:1973ee} we
have used delta-shell potentials for the inner {\it unknown part} to
undertake a simultaneous partial wave analysis (PWA) to proton-proton
and neutron-proton scattering data from $1950$ to $2013$ below pion
production threshold up to LAB energies of $350$
MeV~\cite{Perez:2013mwa} following the pattern of
Fig.~\ref{fig:NN-anatomy} and taking a charge dependent one pion
exchange (OPE) potential above $r_c$ together with electromagnetic
effects, vacuum polarization, magnetic moment effects
\cite{Perez:2013jpa}. The delta-shell potential reduces the numerical
effort tremendously and enables a fast determination of the covariance
matrix whence errors can be determined and propagated for phase-shifts
or nuclear matrix elements.  With a total of $46$ fitting parameters
we obtain $\chi^2/{\rm d.o.f} =1.06$.  The consistent database
selected in \cite{Perez:2013jpa} uses the improved $3\sigma$ criterion
proposed by Gross and Stadler~\cite{Gross:2008ps} which allows to
rescue data which would otherwise have been discarded, see
Fig.~\ref{fig:NN-abundance}. Data and other amusements can be found at
the NN provider Android app, available at Google Play Store, see right
panel of Fig.~\ref{fig:NN-anatomy}.

\begin{table}[t]
\caption{Complete  NN database from PWA without rejection. $N_{\rm Data}=8124$.}
\centering
\label{tab:1}       
	\begin{tabular}{llllllllll}
\hline\noalign{\smallskip}
      $r_c$ [fm]  &          & 1.8 &   &  & 2.4  & & & 3.0 &  \\ 
             &    & $N_{\rm p}$ & $\chi^2/\nu$ &   & $N_{\rm p}$ & $\chi^2/\nu$ &  & $N_{\rm p}$ & $\chi^2/\nu$ \\
\tableheadseprule\noalign{\smallskip}
      OPE  &   & 31   & {1.80}     &  & 39   & 1.56     &  & {46}   & {1.54} \\
      TPE(NLO) &   & 31   & {1.72}     &  & 38   & 1.56     &  & 46   & 1.52 \\
      TPE(N2LO) & \ \ \ \ \ \ \ \ \  & {30+3} & {1.60}  & \ \ \ \ \ \ \ \ \  & 38+3 & 1.56     & \ \ \ \ \ \ \ \ \ & 46+3 & 1.52
	\end{tabular} 
\end{table}
\begin{table}[t]
\caption{$3\sigma$-selected  NN database from potential analysis.}
\centering
\label{tab:2}       
	\begin{tabular}{llllllllll}
\hline\noalign{\smallskip}
      $r_c$ [fm]  &          & 1.8 &   &  & 2.4  & & & 3.0 &  \\ 
             & $N_{\rm accept}$   & $N_{\rm par}$ & $\chi^2/\nu$ &   $N_{\rm accept}$  & $N_{\rm par}$ & $\chi^2/\nu$ &  $N_{\rm accept}$ & $N_{\rm par}$ & $\chi^2/\nu$ \\
\tableheadseprule\noalign{\smallskip}
      OPE  & {5766}  & 31   & 1.10     &    6363 & 39   & 1.09     &  {6438} & {46}   & 1.06 \\
      TPE(NLO) & {5841}  & 31   & 1.10     &    6432 & 38   & 1.10     &    6423 & 46   & 1.06 \\
      TPE(N2LO) & {6220}  & {30+3} & 1.07     &    6439 & 38+3 & 1.10     &    6422 & 46+3 & 1.06
	\end{tabular}
\end{table}
\begin{table}[t]
\caption{Consistent NN database from the improved $3\sigma$-criterion. $N_{\rm Data}= N_{\rm accept}^{({\rm OPE},r_c=3 {\rm fm})}=6713$.}
\centering
\label{tab:3}       
	\begin{tabular}{lllllll}
\hline\noalign{\smallskip}
      $r_c$ [fm]  & 1.8    &       & 2.4 & &  3.0 & \\
        & $N_{\rm par}$ & $\chi^2/\nu$   & $N_{\rm par}$ & $\chi^2/\nu$  & $N_{\rm par}$ & $\chi^2/\nu$ \\
\tableheadseprule\noalign{\smallskip}
      OPE  &  31   & {1.37}   & 39   & 1.09    & {46}   & {1.06} \\
      TPE(NLO)  & 31   & {1.26}      & 38   & 1.08    & 46   & 1.06 \\
      TPE(NNLO) & {30+3} & {1.10}     & 38+3 & 1.08     & 46+3 & 1.06
	\end{tabular}
\end{table}

We have also explored the role of chiral two pion exchange ($\chi$TPE)
interactions at intermediate and long
distances~\cite{Perez:2013oba}. Comparison of OPE and TPE results are
given in tables \ref{tab:1},\ref{tab:2} and \ref{tab:3}.  In table
\ref{tab:1} we show the $\chi^2$ values corresponding to a direct fit
to all the data. These large values prevent error propagation.  In
table \ref{tab:2} we show the $\chi^2$ values corresponding to a
dynamical data base fit to all the data subjected to the $3\sigma$
criterion, so that the selection of the data depends on the potential.
As we see, there is some improvement but data differ.  Finally, in
table \ref{tab:3} we use the fixed and consistent data from the OPE
$r_c=3 {\rm fm}$ analysis. An acceptable $\chi^2=1.1$ with 30
parameters, see Eq.~(\ref{eq:Npar}), allows to propagate errors. In
${\rm GeV}^ {-1}$ units we obtain~\cite{Perez:2013oba}
\begin{eqnarray}
c_1= -0.41 \pm 1.08  \qquad  c_3=-4.66 \pm 0.60  \qquad c_3=4.31 \pm 0.17 
\end{eqnarray}
and a correlation $r(c_1,c_3) =-1 $. This result depends crucially on
making the fit up to $E_{\rm LAB} \le 350 {\rm MeV}$.

\section{Discussion}

One may wonder why should one determine NN interactions by fitting to
higher energies than actually resolved in light nuclei. For instance,
one can fit the $^1S_0$ and $^3S_1$-waves scattering length and
effective ranges with just one attractive
delta-shell~\cite{NavarroPerez:2011fm}, yielding a triton and
$\alpha$-particle binding energies of $(B_t,B_\alpha)=(5.2,20.0){\rm
  MeV}$. Variational mean field shell model calculations yield binding
energies for $^4$He, $^{16}$O and $^{40}$Ca at $20\%$ level when
phases are fitted below LAB energy, $E_{\rm LAB} \le 125 {\rm
  MeV}$. This is so because the interaction becomes soft and short
distance correlations become marginal. When the full amplitude is
fitted in that energy range errors grow dramatically making for
instance $\chi$TPE statistically invisible vs
OPE~\cite{Amaro:2013zka}.  The binding of light nuclei does not depend
explicitly on the high NN scattering data, but the {\it accuracy} of
the interaction does. Predictive power can still be achieved by
solving the many body problem to this
accuracy~\cite{NavarroPerez:2012vr,Perez:2012kt}.


\end{document}